\documentclass[xcolor=dvipsnames]{article}
\usepackage[sort&compress,numbers]{natbib}

\usepackage{geometry}
\usepackage{graphics}
\usepackage{graphicx}

\usepackage{color}

\definecolor{Pergamen}{RGB}{220,210,200}
\definecolor{LightGray}{RGB}{240,240,235}
\definecolor{PaleBlue}{RGB}{190,210,255}
\definecolor{DarkGreen}{RGB}{0,80,20}
\definecolor{SoftRed}{RGB}{255,220,170}
\definecolor{DarkBlue}{RGB}{0,10,80}
\definecolor{Blue}{RGB}{0,20,100}
\definecolor{DarkRed}{RGB}{80,10,10}
\definecolor{Red}{RGB}{160,20,20}

\usepackage[english]{babel}
\usepackage[latin1]{inputenc}

\usepackage{times}
\usepackage[T1]{fontenc}

\usepackage{graphics}
\usepackage{graphicx}

\usepackage{marvosym}
\usepackage{wasysym}
\usepackage{bm}
\usepackage{setspace}

\usepackage{textpos}
\usepackage{hyperref}

\title{%
\bf  Gintropy:\\
 Gini index  based generalization of Entropy}
\author{{\sc T.~S.~Bir\'o$^{1,2}$} \quad and \quad {\sc Z.~N\'eda$^3$ }
\\[1em] $^1$Wigner Research Centre for Physics, Budapest, Hungary; 
\\ $^2$ Complex Science Hub Medical University Vienna, Austria 
\\ $^3$ Babe\c{s}-Bolyai University, Department of Physics, Cluj, Romania
}

\newcommand{\vs}{\vspace{3mm}}

\newcommand{\be}{\begin{equation}}
\newcommand{\ee}[1]{\label{#1} \end{equation}}
\newcommand{\ba}{\begin{eqnarray}}
\newcommand{\ea}[1]{\label{#1} \end{eqnarray}}

\newcommand{\ov}[1]{\overline{#1}}

\newcommand{\ver}[1]{\left\vert #1  \right\vert}
\newcommand{\exv}[1]{ \left\langle {#1} \right\rangle}

\newcommand{\pt}[2]{ \frac{{\rm d} #1}{{\rm d} #2}}

\newcommand{\ead}[1]{ {\rm e}^{#1}}

\newcommand{\infi}{ \int_{0}^{\infty}\limits\!\!}
\newcommand{\inxi}[1]{ \int_{#1}^{\infty}\limits\!\!}
\newcommand{\inzx}[1]{ \int_{0}^{#1}\limits\!\!}
\newcommand{\inab}[2]{ \int_{#1}^{#2}\limits\!\!}

\begin{document}

\maketitle

\begin{abstract}
Entropy is being used in physics, mathematics, informatics and in related areas to
describe equilibration, dissipation, maximal probability states and optimal compression of
information. The Gini index on the other hand is an established measure for
social and economical inequalities in a society. In this paper we explore
the mathematical similarities and connections in these two quantities and introduce 
a new measure that is capable to connect these two at an interesting analogy level. This 
supports the idea that a generalization of the Gibbs--Boltzmann--Shannon
entropy,  based on a transformation of the Lorenz curve, 
can properly serve in quantifying different  aspects of complexity 
in socio- and econo-physics.
\end{abstract}

\section{Introduction}

\subsection{Motivation}
Many researchers use entropy as an appropriate measure for quantifying complexity or 
the inequality level in a complex system. There is an overwhelming choice in generalized entropy formulas, 
some of them satisfying more of the basic axioms than the others \cite{THURNER2017}. 
The classical Boltzmann-Gibbs-Shannon formula 
is often used in economic and social studies without 
elaborating to much on the conditions under which it is an appropriate thermodynamic function. 
Most prominently the additivity of entropy upon the factorization of probabilities is, as a rule, not 
tested and therefore the use of entropy remains at the level of a crude analogy. Using 
the Tsallis- or R\'enyi entropy formula \cite{AMIGO2018} is also not a sufficient choice. Although a free parameter in this entropy provides more flexibility in processing and interpreting statistical data and generalizing the additivity, there is no basic reason why not to use yet another formula that satisfies the basic 
physical requirements for the entropy. 

On the other hand the most popular way for quantifying the inequality level in a socio-economic system is to use the Gini index,
introduced first time by the economist Corrado Gini \cite{GINI1914}. This measure provides a simple method of quantifying the
deviation from a uniform distribution, and it is not a quantity borrowed by a simple analogy from thermodynamics. It also has the advantage that
its value is a number in the $[0,1]$ interval, alike an order parameter. The Gini index is $0$ when all members of  the investigated society
are equal in the relevant quantity and it is $1$ if one member is monopolizing the whole of the available resources. The Gini index can be determined 
experimentally either graphically by constructing the Lorentz curve \cite{ORIGLORENZ1905}, or by the simple formula
\begin{equation}
G=\frac{1}{\langle x \rangle} \frac{\sum_{i=1}^{N}\sum_{j=1}^{N} | x_i-x_j |}{2N^2},
\end{equation} 
where $x_i$ is the relevant quantity for element $i$, and $\langle x \rangle$ is its average value for the whole system with $N$ elements. 
While the Gini index is traditionally
used to measure wealth-, income- or other inequality, 
the entropy is a concept stemming from physics and 
mathematics and is applied to understand, describe and construct optimal or
equlibrium distributions. At the first glance these two termini show no reason
to be connected. However, in recent publications it has been observed that the Gini index and
the total Shannon entropy of socio-economical models and data show a
synergic behavior \cite{MARMANI2020}.

In this paper we shall demonstrate that the mathematical construction formulas of 
the Gini measure of inequality in a society on the one hand and the 
entropy--probability trace formula on the other hand 
bring intriguing similarities at a certain step of their derivation. 
Both quantities are integrated quantities, in the sence
of summing over alternative values of a basic variable, $x$. 
We propose the usage of the phrase {\em ''gintropy''}
in order to express the combination of the Gini index \cite{GINI1914,ATKINSON1970,SHORROCKS1980}
and the entropy, both associated
to a probability density distribution (PDF).

\subsection{Basics}

Let us consider the relevant quantity of the investigated system as a continuous variable $x$. This could be for example, salary, wealth, population etc...
The occurence frequency
of this given value in a huge set of data is described by the normalized probability density
function (PDF): 
\be
 \infi \rho(x) \, dx \: = \: 1.
\ee{PROBNORM}
An approximation to such mathematical PDF-s is given in the praxis by observing
the number of occurences of values in a short bin $[x,x+dx]$ and dividing these
by their sum, the total number:
\be
 \rho(x)= \lim_{\Delta x \rightarrow 0}\frac{N(x,x+\Delta x)}{N_{{\rm tot}}\cdot \Delta x}
\ee{NPERSUMPROB}
with $N_{{\rm tot}}$ the total number of observed data.
In income distributions for example,  $N(x,x+\Delta x)$ is the number of persons having an income
in the $\Delta x$ interval  starting at $x$. The total income is then obtained as
\be
 X_{{\rm tot}} \: = N_{tot} \: \infi x \, \rho(x) \, dx,
\ee{TOTALX}
and the average income is given by
\be
 \exv{x} \: = \: \infi x \, \rho(x) \, dx \: = \: \frac{X_{{\rm tot}}}{N_{{\rm tot}}}.
\ee{AVX}
Both the entropy and the Gini index can be expressed as expectation values
of some functions of $x$ over the PDF $\rho(x)$, the latter we are going to demonstrate
in the present paper.

\vs \noindent
Not only the PDF-s, but frequently the cumulative distributions are in our light-spot. A first reason for this is that 
the experimental shape of the cumulative functions are smoother even in case of a poorer 
statistics. A second reason is, that especially for income distribution and inequality
the total body of ''rich'' is better contrasted to the ''poor''.

\vs \noindent
It is straightforward to construct the quantity 
''the population fraction of richer than $x$'' as
the tail-cumulative integral of the PDF:
\be
 \ov{C}(x) \: = \: \inxi{x} \rho(y) \, dy.
\ee{CUMULC}
A similar cumulative quantity is the wealth accumulated by this richer class,
divided by the average income:
\be
 \ov{F}(x) \: = \: \frac{1}{\exv{x}} \, \inxi{x} y \, \rho(y) \, dy.
\ee{CUMULF}
Trivially one obtains $\ov{C}(0)=1$ and $\ov{F}(0)=1$.

The famous Pareto-law expresses that $p$ fraction of the population
possesses $(1-p)$ fraction of the wealth. In the original statement about
the economy at the end of 19-th century it was $p=0.2$, formulated as the ''80/20'' rule: 
20 percent of the population having 80 percent of the total wealth
~\cite{PARETO,PARETO-SHORT,ABOUT-PARETO}. 
Later also a ''90/20'' rule has been suggested by Dunford
~\cite{DUNFORD}, this looses however the elegant definition of the {\em Pareto point} (see the next paragraph). 
Analyses of national GDP comparisons and wealth distribution
in certain countries often use in the wealthy region a power-law fit, 
$\rho(X)=c x^{-(1+\alpha)}$, calling the parameter $\alpha$ the Pareto-index
~\cite{LEVY,PIKETTY,SINHA,YAKOVENKO}. It is however largely debated where should one consider the cut-off in the distribution curve, over which
the tail is of power-law type.
For a part of the PDF also exponential fits can be done~\cite{EXPONENTIAL}.
As an overall fit to the whole income distribution curve recently it has been shown that a Tsallis--Pareto
cut power-law or some special beta prime distribution works well~\cite{NEDAETAL}.

For a simple division of the system in an upper and lower class the $x_P$ Pareto-point is used, satisfying:
\be
 \ov{C}(x_P) \: = \: p,  \qquad {\rm while}  \qquad  \ov{F}(x_P) \: = \: 1-p.
\ee{PARETOPOINT}
The implicit relation, $x_P(p)$, depends on the underlying PDF, $\rho(x)$.
Since $\ov{C}(0)+\ov{F}(0)=2$ and the general sum is monotonically decreasing, due to
\be
\pt{}{x}\left(\ov{C}(x) + \ov{F}(x) \right)=-\left(1 + \frac{x}{\exv{x}} \right)\rho(x)
\: \le \: 0,
\ee{CPLUSF}
there is always a point $x=x_P$ where $\ov{C}(x_P)+\ov{F}(x_P)=1$.
However, the value $p$ cannot be arbitrary. 

\vs \noindent
As we shall discuss in the next section, the Gini index, $G$, can be expressed in several 
alternative ways: i) as the average of big differences in the data set, ii) 
as a construction using the above cumulative quantities or iii) as an expectation
value of the cumulative of the cumulative. $G$ expressed as an integral over 
$\ov{C}$ contains an integrand $\sigma(\ov{C})$.
For some PDF-s this function turns out to be formally identical with the terms in 
entropy -- probability trace formula known from elsewhere. 
These formulas define the {\em gintropy}, as a function
of the cumulative measure of being ''richer than'', $\sigma(\ov{C})$ -- and
this function coincides with the classical entropy for an exponential PDF, 
alike the Gibbs--Boltzmann distribution of energy in thermodynamics. 
For some other, frequently considered distributions in complex systems
the {\em gintropy} resembles terms of
various generalizations of the Gibbs--Boltzmann-Shannon entropy. Among others
we arrive at the Tsallis-entropy for the original Pareto distribution, and some further interesting cases.
By construction, as we shall demonstrate later, the {\em gintropy} curve is the difference
between the Lorenz curve and the diagonal in the $\ov{F}$ vs $\ov{C}$ maps.

In the sequel of this paper we explore these formulas as several facets of the Gini
index and its calculation. After the mathematical definitions and equivalent forms
we present certain analytically given PDF-s, each reflecting a theoretical
possibility about income inequalities: extreme communism giving every person the
same income; divided society defining two classes of the previous case with a fixed share;
eco-window, providing equal probability to any income in a fixed, but
possibly even infinite interval; the exponentially distributed income taken as
an analogy to the nature of atomic physics; and finally the Pareto-distribution
characteristic to capitalism. To each model a different Gini index, $G$, and
also a different {\em  gintropy}, $\sigma(\ov{C})$ belong.  Finally we collect a few
ideas about what laws the Gini index and {\em gintropy} may follow: is there a trend
akin to the second law of thermodynamics? Are societies closed systems or not?
Can or must inflation distort our analysis?

\section{Gross Inequality in general}

Let $\rho(x)$ be a normalized PDF. The Gini index in the continuous $x$ case is defined as:
\be
 G \: \equiv \: \frac{1}{2\exv{x}} \, \infi dx \, \infi dy \, \ver{x-y} \, \rho(x) \, \rho(y)= \frac{1}{\exv{x}} \, \infi dx \,\int_{x}^{\infty}\limits\!\! dy \, (y-x) \, \rho(x) \, \rho(y).
\ee{GINIDEF}
It can easily be proven that its value is always between zero and one, and is used
to quantify the gross inequality in the distribution $\rho(x)$.
The original definition (\ref{GINIDEF}) can be expressed by using the cumulatives as
\be
  G \: = \: \infi \, \rho(x) \, \left[ \ov{F}(x) \, - \, \frac{x}{\exv{x}} \, \ov{C}(x) \right] dx.
\ee{GINDEF2}
This expression can be further comprised by considering the cumulative of the cumulative: 
\be
	\ov{h}(x) \: \equiv \: \inxi{x} dy \, \ov{C}(y) \: = \: \inxi{x} dy \inxi{y} dz \, \rho(z)
	\: = \: \inxi{x} dz \inab{x}{z} dy \, \rho(z) \: = \: \inxi{x}  (z-x) \rho(z)  \, dz=\exv{x} \ov{F}(x)-x\ov{C}(x).
\ee{CUMCUMDEF}
Finally, from here  the Gini index is then expressed as a ratio of two expectation values:
\be
 G \: = \: \frac{1}{\exv{x}} \infi \rho(x) \, \overline{h}(x) \,  dx
   \: = \: \frac{\exv{\ov{h}(x)}}{\exv{x}}.
\ee{GINDEF3}
Alternatively it can be {expressed via the cumulative population}, solely.
From the corresponding definitions we have the derivatives: 
$\rho(x)=-d\ov{C}/dx$, $x\rho(x)=-\exv{x}d\ov{F}/dx$ and 
therefore $x=\exv{x}d\ov{F}/d\ov{C}$. 
Using the immediate equations $\rho(x)=\pt{^2\ov{h}(x)}{x^2}$ and $\ov{C}(x)=-\pt{\ov{h}(x)}{x}$, 
from eq. (\ref{GINDEF3}), and integrating by parts we get:
\be
 \exv{x} \, G \: = \: \infi  \ov{h} \, \pt{^2\ov{h}}{x^2} \, dx\: = \:
	\ov{h}(0)\ov{C}(0) - \infi  \ov{C}^2(x) \, dx.
\ee{GINIBYPARTINT}

Using the boundary conditions $\ov{C}(0)=1$ and $\ov{h}(0)=\exv{x}$ we arrive at
\be
	G \: = \: 1 \, - \, \frac{1}{\exv{x}}\infi  \ov{C}^2(x) \, dx
	\: = \: \frac{1}{\exv{x}} \infi  \ov{C} \, (1-\ov{C}) \, dx.
\ee{GINFORMC}
This form reminds to the quantum impurity measure, $Tr(\rho-\rho^2)$, which is zero
only for pure states. In the theory of searching trees in informatics, the expression
$I_G=\sum_i(p_i-p_i^2)$ is called Gini impurity measure~\cite{GINIINDEX}.
Let us also note here that for scaling PDF-s, 
i.e $\rho(x)=\frac{1}{\exv{x}} \, f\left(\frac{x}{\exv{x}}\right)$, the cumulative
functions, $\ov{C}$, and the Gini index, $G$, do not depend directly on $\exv{x}$, it depends only on the
form of the $f(z)$ function.
This is important when studying the history (time evolution) of $G$ and
the related constructions: an overall inflation increasing $\exv{x}$
in time, will not influence this inequality measure.

\begin{figure}[bh]
	\includegraphics[width=0.55\textwidth]{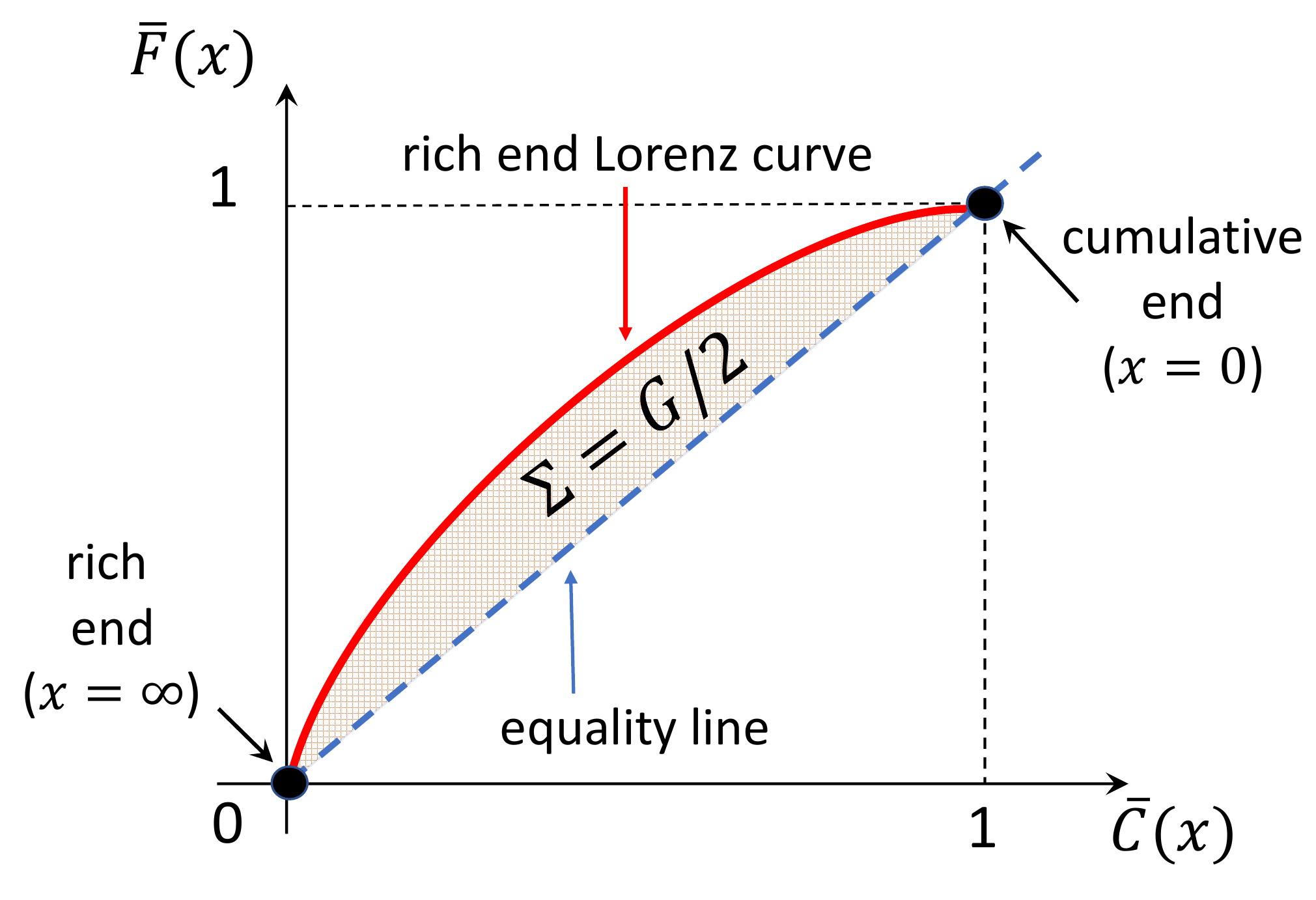}
	\includegraphics[width=0.45\textwidth]{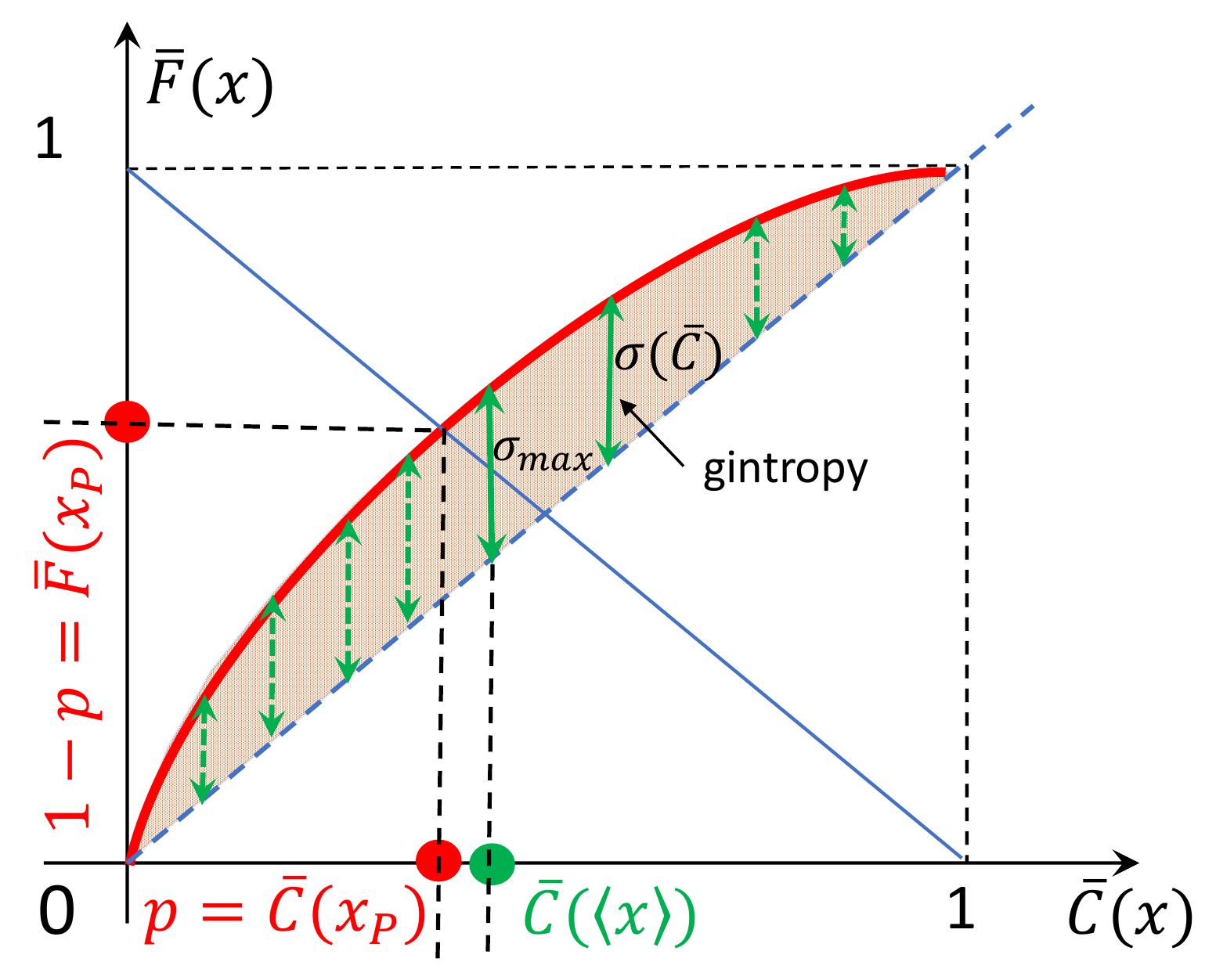}
	
\caption{ \label{Lorenz-curve}
	(left) The rich end Lorenz curve, and connection with the Gini index. (right) 
	Visual illustration of the gintropy, the Pareto Point and the maximal gintropy. }  
	
\end{figure}

\vs \noindent
Finally we arrive now at the construction of the quantity {\em gintropy}. A fashionable
representation of the Gini index is realized by plotting the cumulative
wealth percentage in terms of the cumulative population possessing that wealth like in Figure \ref{Lorenz-curve}.
It can be shown that the half-moon area between the Lorenz curve
~\cite{ORIGLORENZ1905,IRITANI1983,THISTLE1989,AABERGE2000,LORENZCURVE2005} and the diagonal 
of the unit square (known as {\em equality line}) in such an $\ov{F}(x)$ vs $\ov{C}(x)$ plot, 
\be	
\Sigma \: \equiv \: \inab{0}{1} \sigma(\ov{C}) \, d\ov{C},
\ee{SIGMADEF}
is exactly $G/2$. 
The integrand, $\sigma(\ov{C})$, under the integral over $\ov{C}$ 
-- which runs between zero and one 
-- behaves alike an 
entropy-density\footnote{The original and nowadays used Lorenz curve actually
maps the low-cumulatives, integrated from zero to $x$. However, $\sigma(C)$ instead
of $\sigma(\ov{C})$ does not remind to entropy formulas.}. 
We call this quantity {\bf  gintropy}, and define as the difference between
the rich-end-cumulative Lorenz curve and the diagonal: 
\be
 \sigma(x) \: \equiv \: \ov{F}(x) \, - \, \ov{C}(x) \: = \: 
 \inxi{x} \left(\frac{y}{\exv{x}} - 1 \right) \, \rho(y) dy.
\ee{GINTRODEF}

From the above definition $\sigma(\ov{C})$ remains to be reconstructed with the
help of $\ov{C}(x)$. We note that using the relations, $\ov{C}(x)=1-C(x)$ and
$\ov{F}(x)=1-F(x)$ this quantity equivalently can be expressed by the poor-end-cumulative
Lorenz curve (the generally used form of the Lorenz curve) , too:
\be
 \sigma(x) \: = \: C(x) \, - \, F(x) \: = \:
 \inzx{x} \left(1 - \frac{y}{\exv{x}} \right) \, \rho(y) dy.
\ee{GINBTROPOOR}
The Gini index is expressed from the gintropy as a simple integral
\be
G \: = \: 2 \, \Sigma \: = \: 2 \inab{0}{1} \: \sigma(\ov{C})  \, d\ov{C}
\: = \: 2 \, \infi \sigma(x) \rho(x) \, dx \: = \: 2 \exv{\sigma(x)}.
\ee{GASTWOSIGMA}
We note here that for any integral one substitutes
$\inab{0}{1} f(\ov{C}) \, d\ov{C}  \: = \: \infi f(\ov{C}(x)) \,  \rho(x) \, dx$.

It is interesting to summarize the proof of this statement here, because
it is a central motivation of thinking in terms of gintropy. Using the respective definitions of
the tail-cumulative quantities, the half-moon area (\ref{SIGMADEF}) is calculated as the
following double integral:
\be
\Sigma \: = \: \infi dx \, \rho(x) \, \inxi{x} dy \, \left( \frac{y}{\exv{x}} - 1\right)
	\, \rho(y) .
\ee{SIGDEF}
Changing the order of integration  leads to
\be
\Sigma \: = \: \infi dy \inab{0}{y} dx \, \rho(x) \left(\frac{y}{\exv{x}}-1 \right) \, \rho(y)
	\: = \: \infi dy \, \left(1-\ov{C}(y)\right) \left(\frac{y}{\exv{x}}-1\right) \, \rho(y).
\ee{PROOFSIGMA}
Here the term with $1$ in the first parenthesis integrates to zero due to the definition of the
expectation value, $\exv{x}$. Then we replace $-\ov{C}(y)\rho(y)=\frac{1}{2}\pt{}{y}\ov{C}^2$,
integrate by parts and compare the result to (\ref{GINFORMC}) to conclude:
\be
\Sigma \: = \: \ \frac{1}{2} \ov{C}^2(0) - \frac{1}{2\exv{x}} 
  \infi dy \, \ov{C}^2(y) \: = \: \frac{1}{2} \, G.
\ee{PROOFSIGPARTIAL}

Now we explore some basic properties of gintropy. Some of these provides further evidences to consider 
gintropy alike a generalized entropy density.
\begin{enumerate}

\item The gintropy is never negative: $\sigma = \ov{F}-\ov{C}\ge 0$
	is proven by inspecting the integral
		$$ 
		\sigma(x) \: = \: \inxi{x} (y/\exv{x}-1)\rho(y) dy 
		\: = \: \inab{0}{x} (1-y/\exv{x}) \rho(y) dy \: \ge \: 0,    
		$$
	and taking the first form for $x \ge \exv{x}$, the second form for the opposite case. 
	This implies that the rich-end wealth fraction is always bigger or equal to
	the population fraction possessing it. 

\item The gintropy is maximal at $x=\exv{x}$, $\sigma_{max}=\sigma(\langle x \rangle)$, since  $d\sigma/dx=(1-x/\exv{x})\rho(x)$ changes 
	its sign exactly there and only there.

\item	According to eq.(\ref{PARETOPOINT}) at the Pareto-point the gintropy equals to 
	$\sigma(x_P)=1-2p$, and therefore for the Pareto point $p \le 1/2$ holds for the rich fraction. 
        Since $\sigma_{max}\ge \sigma(x_P)$, in order to get a 
	Pareto point: $\sigma(\langle x \rangle)\ge 1-2p$, i.e. the maximum of the
	gintropy has to be bigger than this difference value. As a consequence
	for the Pareto Point we have a restriction imposed by the maximal gintropy 
	$(1-\sigma(\exv{x}))/2 \le p \le 1/2$.

\item	The expectation value of gintropy is the half of the gini index:
	$\infi \sigma(x) \, \rho(x) \, dx \: = \: \inab{0}{1} \sigma(\ov{C}) \, d\ov{C} \: = \: \Sigma \:= \: G/2$.

\item	The integral of gintropy over the base value $x$ is the non-Poissonity index,
	$\infi \sigma(x) \, dx  = \frac{{\rm Var}(x)}{\exv{x}}$,  with
	${\rm Var}(x) = \exv{x^2}-\exv{x}^2$ being the variance of $x$. 
	The proof of this statement uses 
	the same mathematical trick as the one in eq. (\ref{CUMCUMDEF}).

\item 	For some particular PDF-s $\sigma(\ov{C})$ looks like an entropy density formula,
	$s(p_i)$.  We present important examples in the next section.

\end{enumerate}

%%%%%%%%%%%%%%%%%%%%%%%%%%%%%
\section{Important Examples}

In this section we list some important examples of the gintropy, $\sigma(\ov{C})$.
We go through primitive models of income/wealth distributions, labelled
as communism, comunism++, eco-window, natural, or capitalism. Starting from model
PDF-s the gintropy expression and the Gini index are calculated.

{\bf Communism}:

Our first example is communism: all incomes are equal, the PDF is simply a
singular delta-distribution, peaked at the single value $a$:
$\rho(x) \: = \: \delta(x-a)$ leading to $\exv{x}=a$, $\ov{C}(x)=\Theta(a-x)$
and $\ov{h}(x)=(a-x)\Theta(a-x)$, with $\Theta(x)$ the Heaviside step function defined as:
\be
\Theta(x) \: = \: \left\{ \begin{array}{c} 0 \qquad x < 0 \\ \\ \frac{1}{2} \qquad x=0 \\ \\ 1 \qquad (x > 0) \end{array}  \right.
\ee{HEAVISIDE}

This leads to $\exv{x}\ov{F}=\ov{h}+x\ov{C}=a \Theta(a-x)$ and by that
\be
 \sigma(x) \: = \:  \ov{F}(x) \, - \, \ov{C}(x) \: = \: 0,
\ee{COMMUNISTICgintropy}
i.e. to an identically vanishing gintropy. 
As a conseqence also {\bf $\bm{G=0}$}. Here no Pareto-point can be found.

{\bf Communism++}

The next example we present is a slight variation of the previous:
now two peaks in a given ratio constitute the PDF. This belongs to
a two-class-society where {\em all are equal but some of them are more equal}.
The two-peak-PDF, $\rho(x)=w\, \delta(x-a)+(1-w)\,\delta(x-b)$ ($b>a$) , delivers 
$\exv{x}=w\,a+(1-w)\,b$. The $w$ fraction of the population has an income
$a$ and the $(1-w)$ fraction $b$.
The cumulative rich population graph shows two steps, at $a$ and $b$, respectively:
\be
 \ov{C}(x) \: = \: w \,\Theta(a-x) + (1-w) \,\Theta(b-x),
\ee{COMM2cumul}
having the value $1$ for $x \le a$,  $(1-w)$ for $x \in [a,b]$, and $0$ otherwise.
Therefore $C(x)=1-\ov{C}(x)$ is zero for $x \le a$, equals to $w$ in the mid interval
and has the value $1$ otherwise. The Gini index is obtained from this as:
\be
 G \: = \: \frac{1}{\exv{x}} \infi \, \ov{C} (1-\ov{C}) dx
 \: = \:  \frac{1}{\exv{x}} \, (b-a) \, w\,(1-w).
\ee{COMM2GINI}
Expressing the weights, $w=\frac{b-\exv{x}}{b-a}$ and $1-w=\frac{\exv{x}-a}{b-a}$, we obtain
the alternative form
\be
	G \: = \: \frac{(\exv{x}-a)(b-\exv{x})}{(b-a)\exv{x}}.
\ee{COMM2GINI2}
It is worth to note that for $a\rightarrow 0$, i.e. when the lower class has (almost)
zero income, the Gini index, cf. (\ref{COMM2GINI}) tends to $G\rightarrow w$, 
exactly the share of the proletars earning $a\rightarrow 0$ in the population. 
This result is independent of $b$, the income in the upper class.

\begin{figure}[h!]
\begin{center}
	\includegraphics[width=0.3\textwidth]{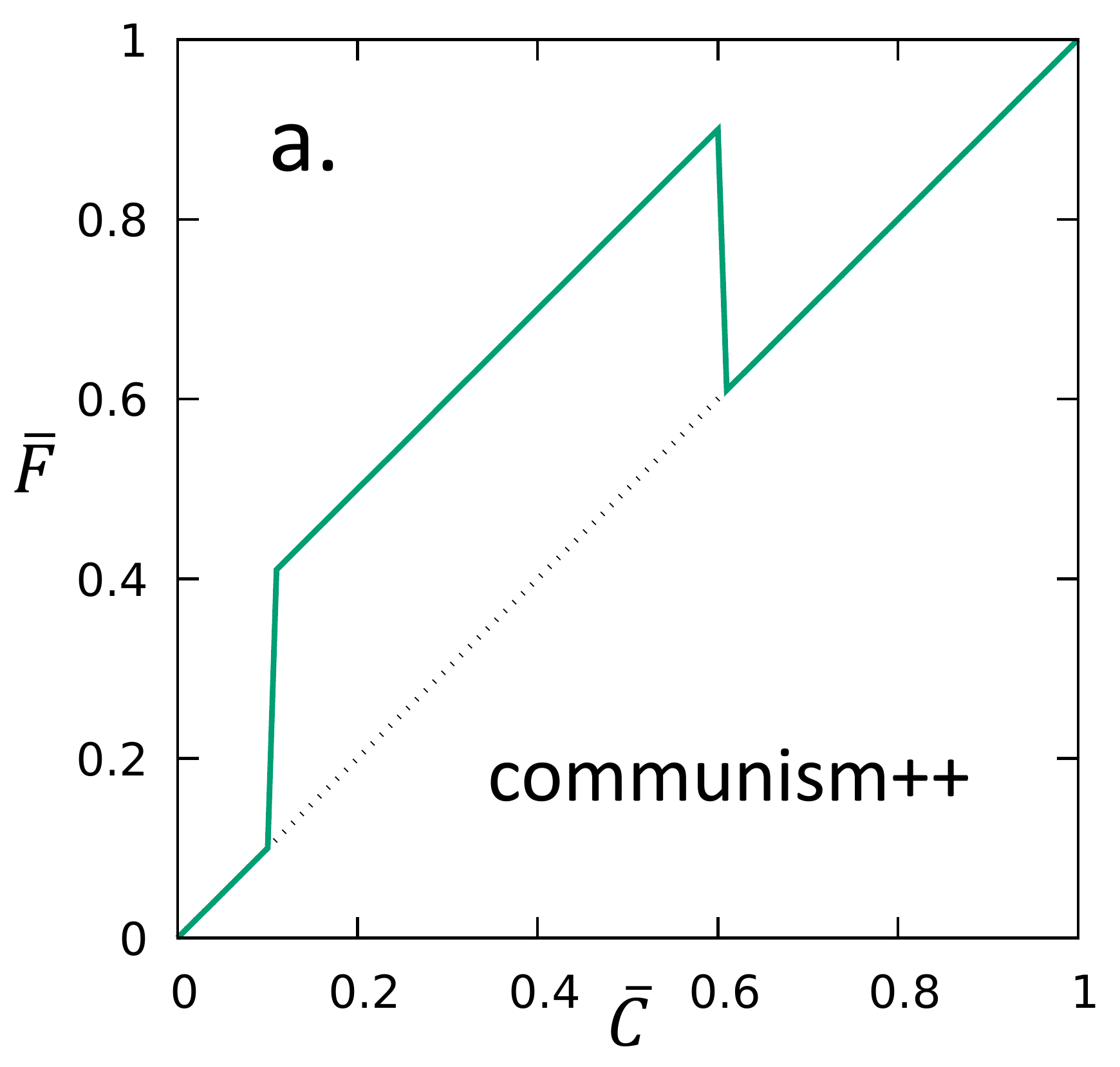} \hspace{1cm}
	\includegraphics[width=0.3\textwidth]{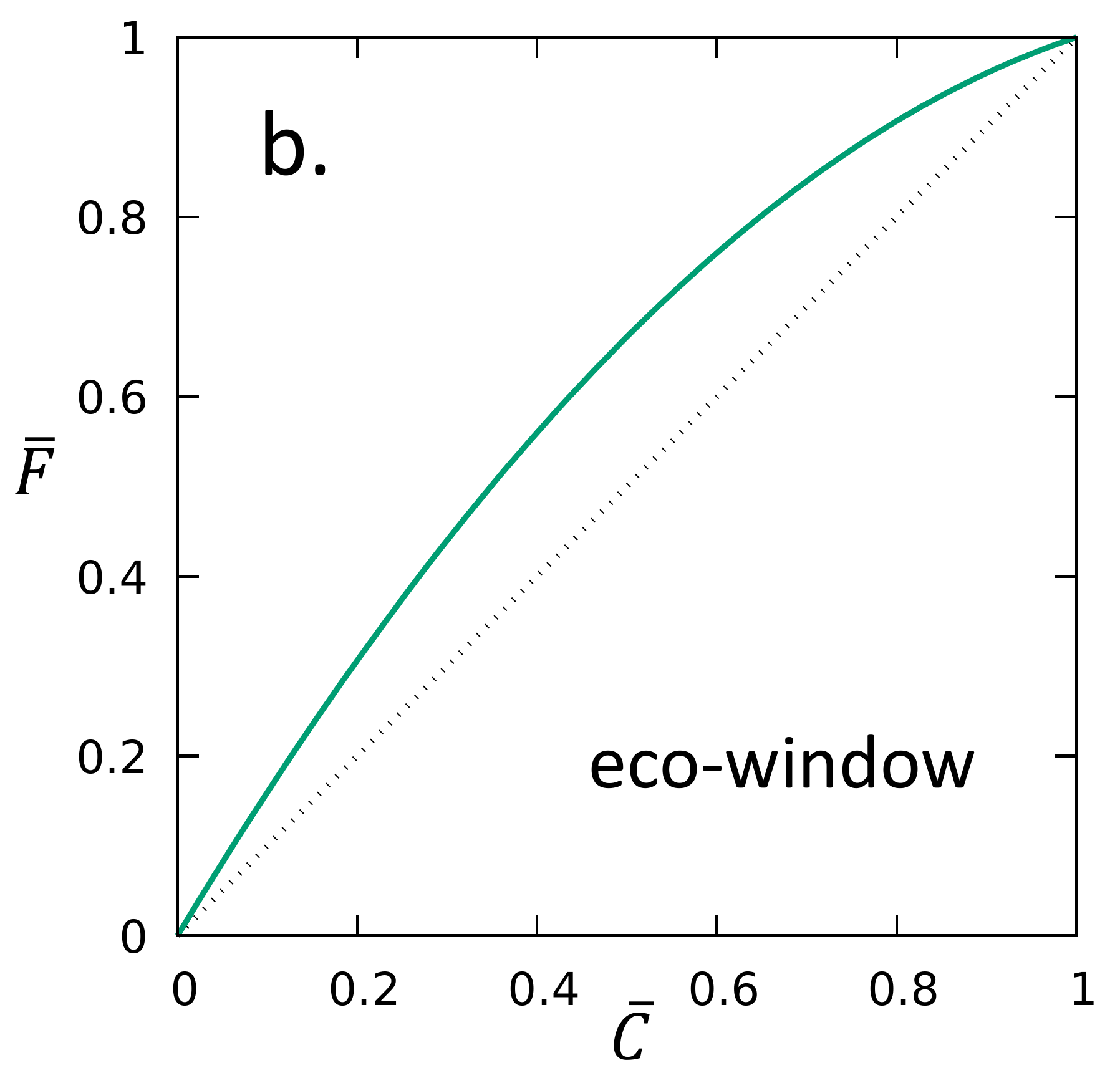}
\vs
	
	\includegraphics[width=0.3\textwidth]{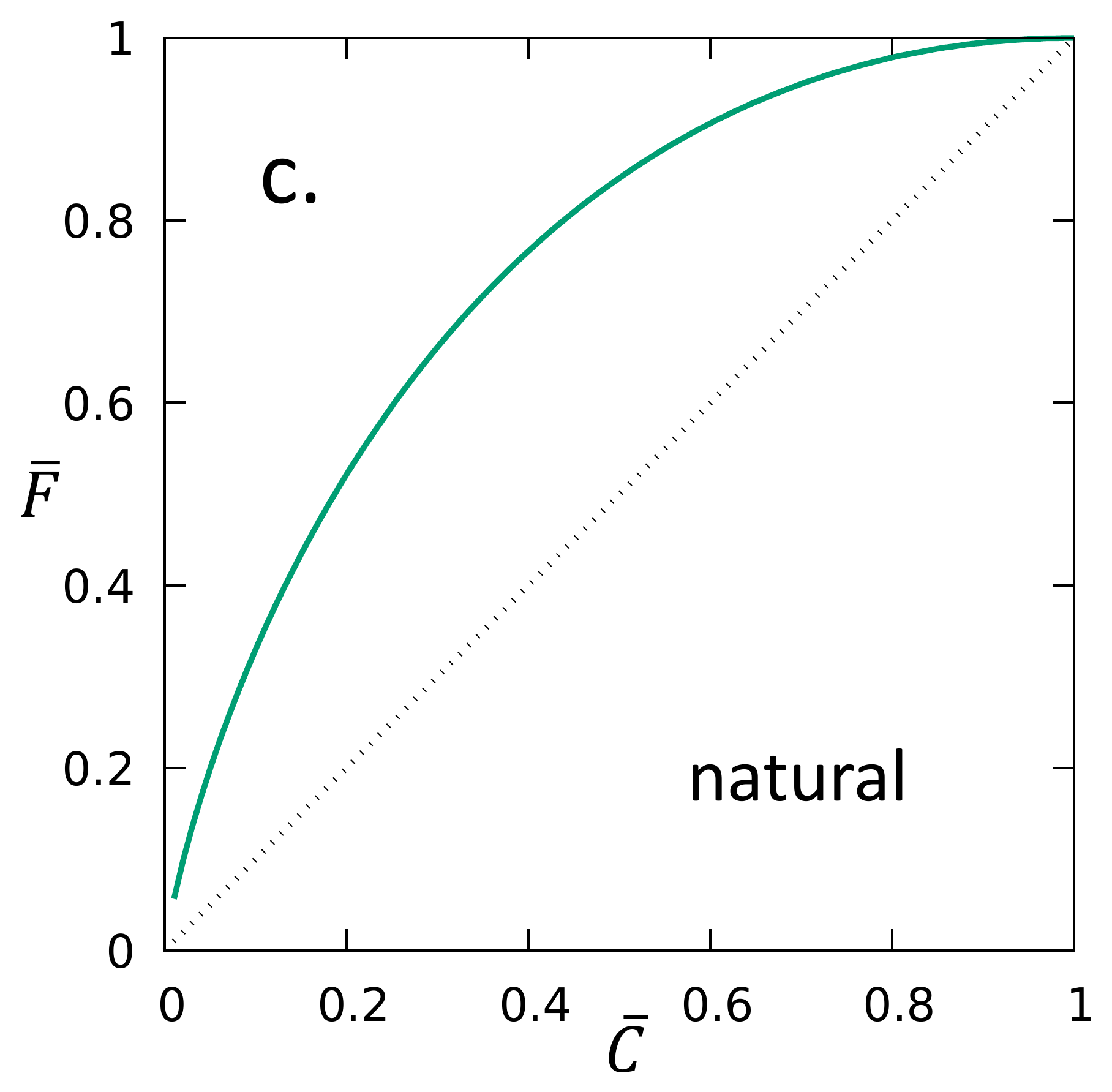} \hspace{1cm}
	\includegraphics[width=0.3\textwidth]{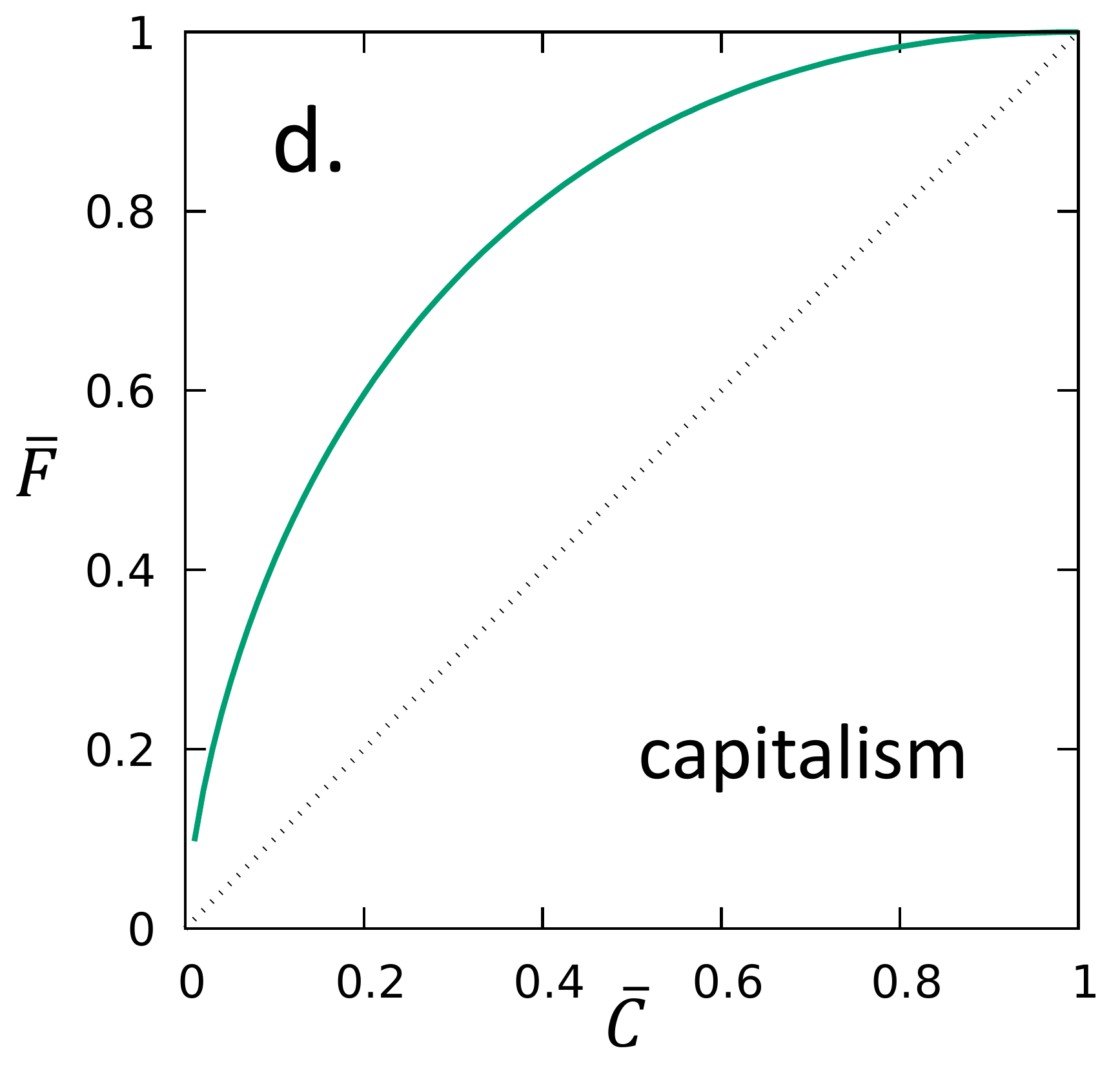}
\end{center}
\caption{ \label{FIG1}
	The $\ov{F}$ vs $\ov{C}$ cumulative maps (Lorenz curves)  for the (a) communism++
	( $a=1$, $b=4$ and $w=0.8$), (b) eco-window ($a=1, b=5$), natural exponential ($\exv{x}=1$)
	and for the capitalism ($A=1, B=3$ $\rightarrow$ $q=3/4$) distributions. 
	The corresponding Gini index are $G=0.3$, $G=2/9$, $G=1/2$ and $G=4/7$, respectively.
}
\end{figure}

The gintropy, following its definition, first is expressed as a function of $x$:
\be
 \sigma(x) \: = \:  \ov{F}(x) \, - \, \ov{C}(x) \: = \: w \left(\frac{a}{\exv{x}}-1 \right) \Theta(a-x) 
	   \, + \, (1-w) \left(\frac{b}{\exv{x}}-1 \right) \Theta(b-x).
\ee{COMM2GINTROP}
It is easy to see that outside the interval $[a,b]$ the gintropy is zero. Inside the interval
only the second term survives giving
\be
 \sigma(x) \: = \: G \, \left[ \Theta(b-x)-\Theta(a-x) \right].
\ee{COMM2SIGINTERVAL}
In conclusion $\sigma(\ov{C})$ shows a plateau at $\ov{C}=1-w$ with the value $G$
and its jumps are at $\ov{C}(a)=1-w/2$ and $\ov{C}(b)=(1-w)/2$:
\be
 \sigma(\ov{C}) \: = \: G \, [\Theta(\ov{C}(a)-\ov{C}) -\Theta(\ov{C}(b) - \ov{C}) ].
\ee{COMM2SIGC}
It is easy to check that indeed
\be
 \Sigma \: = \: \inab{0}{1} \sigma(\ov{C}) \, d\ov{C} \: = \:
 G \left[ \ov{C}(a) - \ov{C}(b) \right] \: = \: G/2.
\ee{COMM2HALFMOON}
The corresponding Lorenz curve is illustrated in Figure \ref{FIG1}a.

\vspace{1cm}

{\bf Eco-window}

The next example is still mathematically simple with a
window-form PDF. We label this as {\em eco-window}: 
here everyone has the same chance for all of possible incomes between $a$ and $b$.
Eventually $a=0$ and/or $b=\infty$ may be considered, as special cases.
For the PDF  $\rho(x) \: = \: \frac{1}{b-a} \, [\Theta(b-x)-\Theta(a-x)] $
one obtains the following cumulative rich distribution:
\be
\ov{C}(x) \: = \frac{b-x}{b-a}\Theta(b-x)-\frac{a-x}{b-a}\Theta(a-x)\:= \left\{ \begin{array}{c} 1 \qquad (x < a) \\ \\ \frac{b-x}{b-a} \qquad {{\rm x\in [a,b]}} \\ \\ 0 \qquad 
(x > b) \end{array}  \right.
\ee{WINABCUMUL}
Obviously $\exv{x}=(a+b)/2$ and according to eq. \ref{GINFORMC} the Gini index becomes:
\be
G \: = \: \frac{1}{\exv{x}}  \inab{a}{b} \frac{(b-x)(x-a)}{(b-a)^2} dx 
	\: = \: \frac{1}{3} \, \frac{b-a}{b+a}.
\ee{WINABGINI}
After some tedious but straightforward calculation the {\em gintropy} is obtained
as a function of $\ov{C}$: 
\be
\sigma(\ov{C})\: = \: 3 \, G \: \ov{C} (1-\ov{C}).
\ee{ANARCHSIGMA}
For a specific choice of $a$ and $b$ the corresponding Lorenz curve is illustrated in Figure 
\ref{FIG1}b. 

{\bf Natural distribution}\\
Our next example is the {natural} distribution, mimicking
the Boltzmann--Gibbs exponential energy distribution, known from statistical physics.
This is not necessarily an equilibrium distribution, it may also be the
stationary limit of ''growth and resetting'' type 
processes with quantity-independent rates \cite{GROWTHRESET}.
The PDF is a scaling one: $\rho(x) \: = \: \frac{1}{\exv{x}} \ead{-x/\exv{x}}$.
The corresponding tail-cumulative probability, the rich population
is given by
\be
\ov{C}(x) \: = \: \ead{-x/\exv{x}},
\ee{CUMEXP}
and the Gini index becomes
\be
  G \: = \: 1 \, - \, \frac{1}{\exv{x}} \infi \ead{-2x/\exv{x}} dx \: = \: \frac{1}{2}.
\ee{EXPONGINI}
Our gintropy formula is constructed as follows: First we obtain the cumulative
of the cumulative,
\be
 \ov{h} \: = \: \inxi{x} \ead{-y/\exv{x}} \, dy \: = \: \exv{x} \, \ead{-x/\exv{x}}.
\ee{EXPHBAR}
From this it is easy to obtain the wealth share of the rich classes,
$\exv{x}\ov{F}=\ov{h}+x\ov{C}=(x+\exv{x})\ead{-x/\exv{x}}$, and based on this the
gintropy
\be
\sigma(x) = \frac{x}{\exv{x}}\ead{-x/\exv{x}}.
\ee{SIGEXPOFX}
In order to express it as a function of $\ov{C}$ we invert (\ref{CUMEXP}) to have
\be
 x(\ov{C}) \: = \: - \exv{x} \, \ln \ov{C}.
\ee{EXPXOFC}
Finally it leads to
\be
\sigma(\ov{C}) \: = \: - \ov{C} \, \ln \ov{C}.
\ee{EXPGINTROPY}
Apart from a constant proportionality factor, this formula formally coincides with 
the terms in the sum of the Boltzmann--Gibbs--Shannon entropy:
\be
S=-k\sum_{i} p_i \ln(p_i)
\ee{S-ENTROPY}
To continue the analogy also $\ov{C} \in [0,1]$. Indeed in this case gintropy
is alike of the entropy density, with the caveat that the cumulative values $\ov{C}(x)$
are never disjunct for different $x$-s, they rather overlap and show a definite
hierarchy. The Lorenz curve for $\langle x \rangle=1$ is illustrated in Figure \ref{FIG1}c. \\

{\bf Capitalism}\\
Our last example is {capitalism}, conjecturing
the base PDF being the cut Pareto (known also as Tsallis-Pareto or Lomax II) distribution~\cite{DOGUM1977}:
\be
\rho(x)=A(B+1)(1+Ax)^{-B-2}
\ee{T-P}
This distribution can also
be obtained as the canonical equilibrium optimizer of the Tsallis entropy~\cite{TSALLIS1988}.
The tail-cumulative integral is
\be
\ov{C}(x)=(1+Ax)^{-B-1},
\ee{TPCUM}
which upon integration leads to the following cumulative of the
cumulative:
\be
\ov{h}(x)=\frac{1}{AB}(1+Ax)^{-B}. 
\ee{TPCUMCUM}	
This result also delivers the expectation value, $\exv{x}=\ov{h}(0)=1/AB$.
The Gini index is calculated in the (\ref{GINFORMC}) form, and it becomes
\be
	G \: = \: 1 - AB \infi (1+Ax)^{-2B-2} dx \: = \: \frac{B+1}{2B+1}.
\ee{TSALLISGINI}
The gintropy as a function of the income, $x$, follows the form
\be
\sigma(x) \: = \:  A(B+1) \, x \, (1+Ax)^{-B-1}. 
\ee{TSALLISGINTROPY}
In order to express this result akin to the entropy, we write $\sigma$
as a function of $\ov{C}$ using the inversion of eq.(\ref{TPCUM})
\be
 x(\ov{C}) \: = \: \frac{1}{A} \left(\ov{C}^{\, -\frac{1}{B+1}} \, - \, 1  \right),
\ee{TPXOFC}
and we obtain
\be
\sigma(\ov{C}) \: = \:	(B+1) \, \left( \ov{C}^{\frac{B}{B+1}} \, - \, \ov{C} \right).
\ee{TPGINTROPOFC}
Finally, using the Tsallis parameter, $q=B/(B+1)$, we arrive at the 
formula:
\be
	\sigma(\ov{C})=\frac{1}{1-q}(\ov{C}^q - \ov{C}),
\ee{TPGIN}
One immediately makes analogy with the terms in the Tsallis entropy formula:
\be
S_q=\frac{k}{1-q}\sum_i (p_i^q-p_i),
\ee{S-TSALLIS}
The Gini index is simply
\be
 G \: = \: \frac{1}{q+1}.
\ee{TPGINIWITHQ}

\begin{figure}[h!]
\centerline{
 \includegraphics[width=0.35\textwidth]{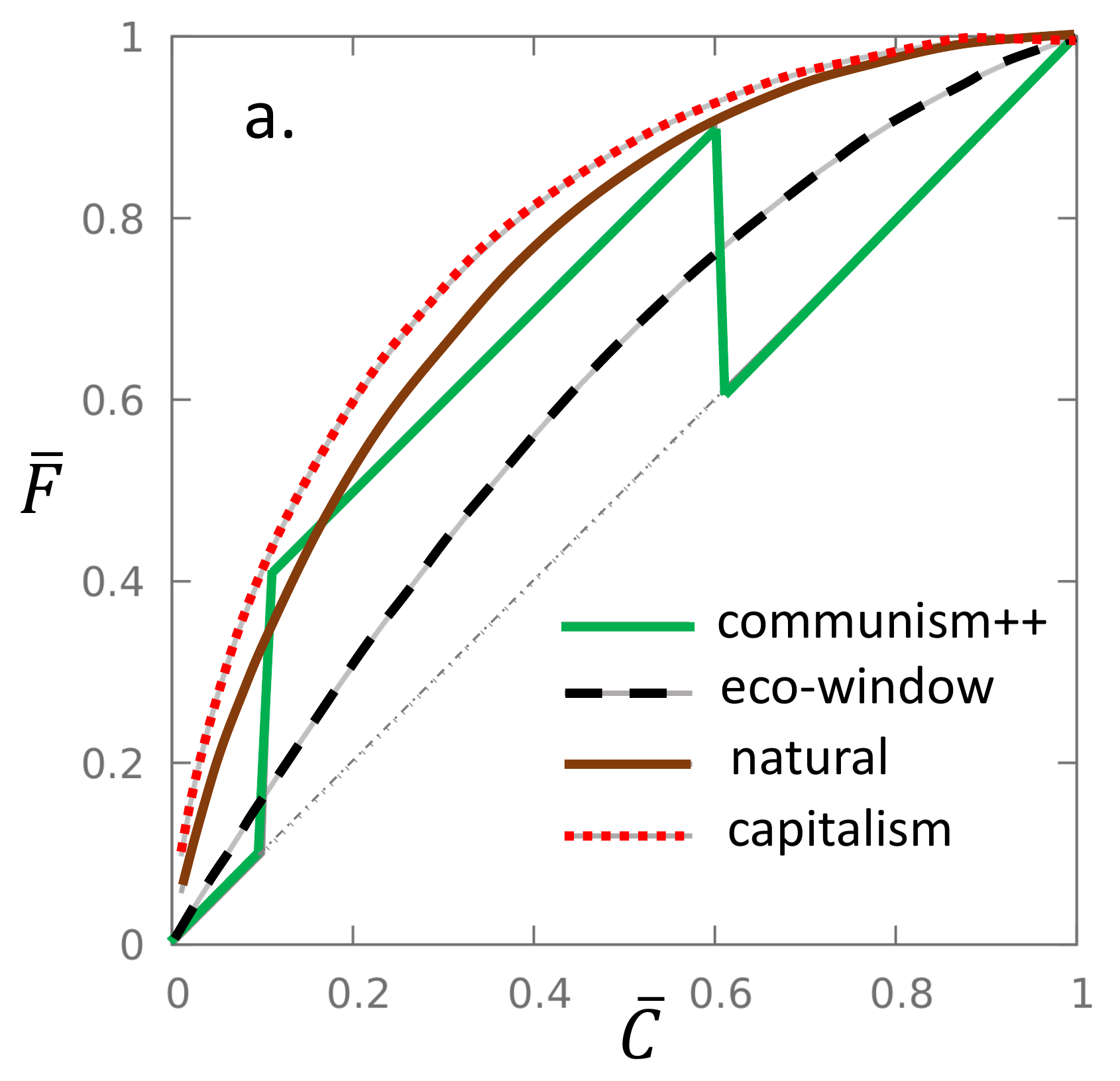} \hspace{1cm}
 \includegraphics[width=0.365\textwidth]{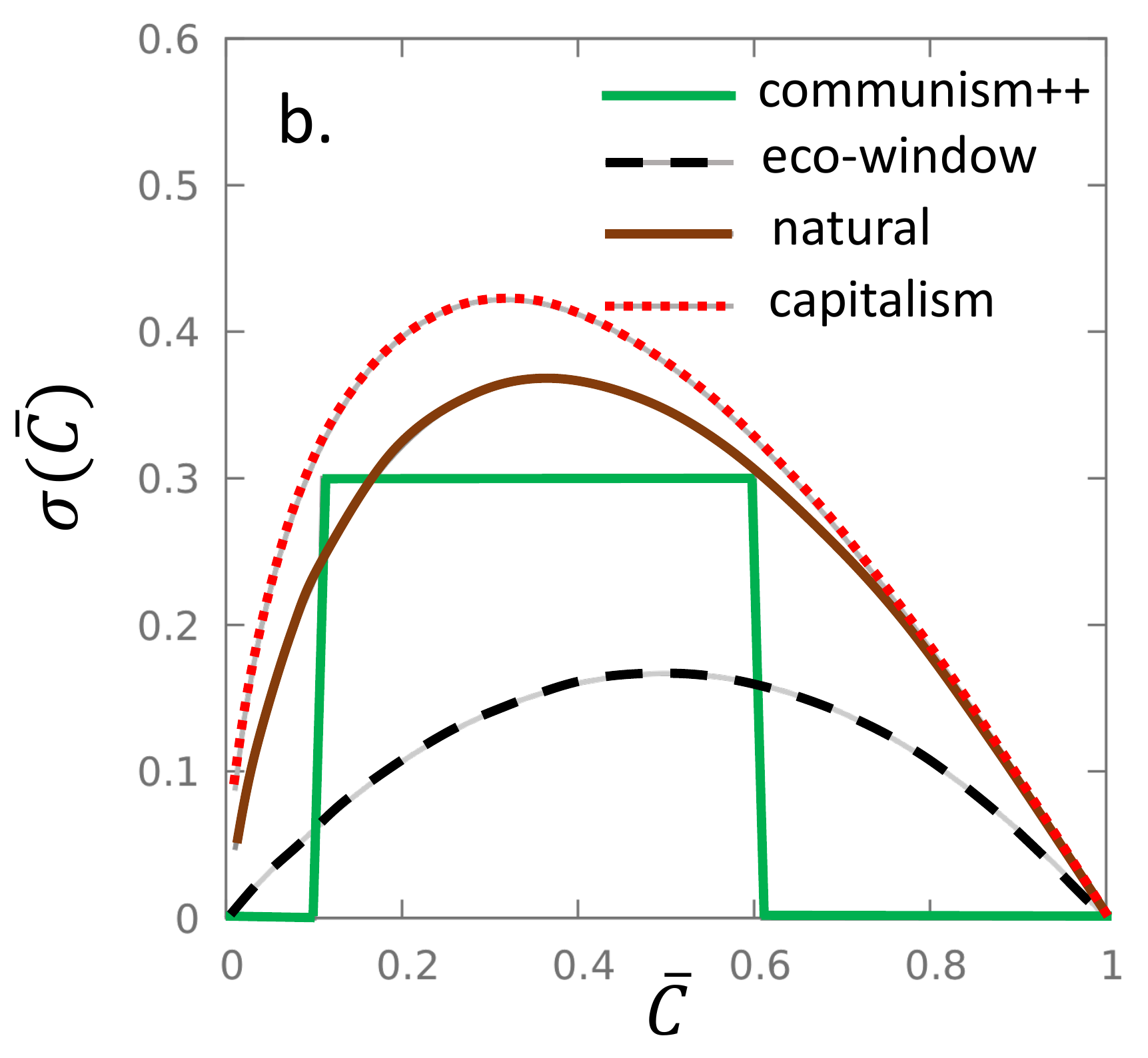}
}
\caption{\label{FIG2}
	(a) $\ov{F}$ -- $\ov{C}$ Lorenz curves in one comparison
	 and (b) the corresponding {\bf gintropy} curves, $\sigma(\ov{C})$,
	for the communism++, eco-window, natural and capitalism models. 
	The Gini indices are $G=0.3$, $G=2/9$, $G=1/2$ and $G=4/7$, respectively.
}
\end{figure}

Similarly with the previously considered cases we illustrate the Lorenz curve for this distributions as well. 
For $A=1$ and $B=3$ the corresponding Lorenz curve is plotted in Figure \ref{FIG1}d.

Finally, we summarise the lesson of the considered theoretical examples in Table \ref{TAB1} and Figure
\ref{FIG2}.

\begin{table}[h!]
\begin{center}
\begin{tabular}{|c|c|c|c|c|c|}
\hline
 & $\rho(x)$ & $\sigma(\ov{C})$ & G \\
\hline 
	natural & $\frac{1}{\exv{x}} \ead{-x/\exv{x}} $ & $-\ov{C} \ln \ov{C}$ & 
	$\frac{1}{2}$ \\
\hline
	capitalism & $\frac{A}{1-q}(1+Ax)^{\frac{-1}{1-q}}$ & $\frac{1}{1-q}\left(\ov{C}^q-\ov{C}\right)$ &
	$\frac{1}{q+1} \ge \frac{1}{2}$ \\
\hline 
	eco-window &  $\frac{1}{b-a}[\Theta(b-x)-\Theta(a-x)]$ &  $3G\, \ov{C} (1-\ov{C})$ &
	$\frac{1}{3} \, \frac{b-a}{b+a} \le \frac{1}{3}$ \\
\hline 
	communism++ & $w\, \delta(x-a)+(1-w)\,\delta(x-b)$ &   $G \, [\Theta(\ov{C}(a)-\ov{C})-\Theta(\ov{C}(b)-\ov{C}) ]$ &
	$\frac{(b-a)q\, (1-w)}{w\,a+(1-w)\,b}$ \\
\hline
	communism & $\delta(x-a)$ & $0$ & $0$ \\
\hline 
\end{tabular}
\end{center}

\caption{\label{TAB1} Summary of PDF-s, the gintropy formulas and Gini index
 values for some ideal income/wealth distribution schemes. 
}
\end{table}

\section{Conclusion}

In this work we  explored a density-like quantity called 
{\em gintropy} which occurs in calculating the Gini index, $G$, for
a given relevant socio-economic distribution, $\rho(x)$. 
This gintropy can be deduced from two cumulative functions,
the rich population fraction and the corresponding richness fraction, $\ov{C}(x)$
and $\ov{F}(x)$, respectively. The proposed ''gintropy'' name, is meant to suggests a connection 
between the inequality measure  quantified by the Gini index and the entropy.    
Its dependence on the rich population fraction cumulative function
reminds to terms in entropy formulas, known from physics, statistics
and informatics. More precisely we found that for the the natural, exponential PDF,
the gintropy reminds 
to the classical Boltzmann--Gibbs--Shannon formula,
$\sigma(\ov{C})=-\ov{C}\ln\ov{C}$. The Gini index is then the
expectation value of the gintropy function, for the exponential PDF
its value is $1/2$. For the Tsallis--Pareto distribution
the Gini index must be always over this value.

Several other PDF-s have been suggested to describe income or wealth distributions
in due of time~\cite{NEDAETAL,MCDONALD1984,TAILLIE1981,WILFLING1993,BOUCHAUD2000}. 
Many of them are not treatable analytically, so
the $\sigma(\ov{C})$ relation can only be explored numerically.

Beyond igniting the theoretical phantasy, the gintropy -- reminding to
generalized entropy formulas -- is also the one-variable density, which
lays under the Gini index, originally defined for measuring inequality.
Turning this statement around, should we seek for such generalizations of
the classical entropy formula which are inequality or impurity measures
at the same time? We beleive that this criterion selects out a subclass
of possible statistical theories among all possible approaches to the
origin, behavior and future of social and economical 
inequalities. Even generalizations of the Gini index formula has been suggested
a few times, cf.~\cite{YITZHAKI1983,KLEIBER2002}. We do not expect
that a corresponding gintropy (''Lorenz curve minus the diagonal'')
would resemble any known entropy formula -- but this question needs further study.

Finally it seems that the ''correct'' entropy measure for economical
and social theories hardly can be a simple copy of the classical formula
known from physics, mathematics and informatics. Our procedure, described
above, is more promising: a recipe for constructing gintropy from
cumulative functions of the underlying PDF whose expectation value is the
half Gini index and whose dependency on the cumulative rich population
coincides with various generalizations of the entropy--probability
formula.

\section*{Acknowledgement}

The work was supported by the Hungarian National Bureau for Innovation,
Development and Research under the project Nr K 123815 and 
 by the UEFISCDI grant: PN-III-P4-ID-PCCF-2016-0084.
T.~S.~Bir\'o thanks for the UBB Star fellowship at the Babe\c{s}-Bolyai
University in Cluj.


\begin{thebibliography}{xxxx}

\bibitem{THURNER2017}
%	{\color{blue} THURNER2017}

S.~Thurner, R.~Hanel and B.~Corominas-Murtra . 
{\em The three faces of entropy for complex systems- information, thermodynamics and the maxent principle}, Physical Review E, {\bf  96} (2017) 032124

\bibitem{AMIGO2018}
%	{\color{blue} AMIGO2018}

J.M.~Amigo, S.G.~Balogh and S.~Hernandez  
{\em A Brief review of Generalized Entropies}, Entropy  {\bf 22} (2020) e20110813


\bibitem{GINI1914}
%	{\color{blue} GINI1914}

C.~Gini:
{\em Sulla misura della concentrazione e della variabilit\'a dei caratteri},
Lettere e Arti {\bf 73} (1914) 1203-1248

\bibitem{ORIGLORENZ1905}
%	{\color{blue} ORIGLORENZ1905}

Max Otto Lorenz:
{\em Methods of measuring the concentration of wealth},
Publications of American Statistical Association,
Vol. 9. (New Series No 70),
p 209-219; 1905.
% original publication establishing the Lorenz curve

\bibitem{MARMANI2020}
%	{\color{blue} MARMANI2020}

S.~Marmani, V.~Ficcadenti, P.~Kaur and G.~Dhesi  
{\em Entropic analysis of votes expressed in Italian
elections between 1948 and 2018}, Entropy  {\bf 22} (2020) e22050523


\bibitem{ATKINSON1970}
%	{\color{blue} ATKINSON1970}

A.~B.~Atkinson:
{\em On the measurement of inequality},
Journal of Economic Theory {\bf 2} (1970) 244-263.

\bibitem{SHORROCKS1980}
%	{\color{blue} SHORROCKS1980}

A.~F.~Shorrocks:
{\em The Class of Additivity Decomposable Inequality Measures},
Econometrica {\bf 48} (1980) 613-625


\bibitem{PARETO}
%	{\color{blue} PARETO}

Vilfredo Pareto:
{\em Cours d' economie politique},
F.Rouge, Lausanne, 1896.

\bibitem{PARETO-SHORT}
%	{\color{blue} PARETO-SHORT}

Vilfredo Pareto:
{\em The New Theories of economics},
Journal of Political Economics (1896) pp 485-502.

\bibitem{ABOUT-PARETO}
%	{\color{blue} ABOUT-PARETO}

Joseph A.~Schumpeter:
		{\em Vilfredo Pareto (1848 - 1923)},
The Quaterly Journal of Economics {\bf 63} (1949) 147-173.

\bibitem{DUNFORD}
%	{\color{blue} DUNFROD}

R.~Dunford, Q.~Su, E.~Tamang et.al.
{\em The Pareto Principle},
The Plymouth Student Scientist {\bf 7} (2014) 140-148.

\bibitem{LEVY}
%	{\color{blue} LEVY}

M.~Levy, S.~Solomon:
{\em New evidence for the power-law distribution of wealth},
Physica A: Statistical Mechanics and its Application {\bf 242} (1997) 90-94.

\bibitem{PIKETTY}
%	{\color{blue} PIKETTY}

I.~Piketty:
{\em Capital in the Twenty-First Century},
Beknap Press 2014.

\bibitem{SINHA}
%	{\color{blue} SINHA}

S.~Sinha:
{\em Evidence for the Power-law tail of the wealth-distribution in  India},
Physica A: Statistical Mechanics and its Application {\bf 359} (2006) 555-562.

\bibitem{YAKOVENKO}
%	{\color{blue} YAKOVENKO}

A.~Dragulescu and V.~M.~Yakovenko:
{\em Statistical mechanics of money},
The European Physical Journal B {\bf 17} (2000) 723-729.

\bibitem{EXPONENTIAL}
%	{\color{blue} EXPONENTIAL}

A.~Dragulescu, V.~M.~Yakovenko:
{\em Exponential and power-law probability distributions of wealth and income
in the United Kingdom and the United States},
Physica A: Statistical Mechanics and its Applications {\bf 299} (2001) 213-221.

\bibitem{NEDAETAL}
%	{\color{blue} NEDAETAL}

Z.~Neda, I.~Gere, T.~S.~Biro, G.~Toth, N.~Derzsy:
{\em Scaling in income inequalities and its dynamical origin},
Physica A: Statistical Mechanics and its Applications {\bf 549} (2020) 124491

%%%%%%%%%%%% ON gini index: impurity %%%%%%%%%
\bibitem{GINIINDEX}
%	{\color{blue} GINIINDEX}

Rekha Molala:
{\em Entropy, Information Gain, Gini Index -- The Crux of a Decision Tree}
https://blog.clairvoyantsoft.com

%%%%%%%%%%%%%%%%%% ON THE LORENZ CURVE %%%%%%%%%%%%

\bibitem{IRITANI1983}
%	{\color{blue} IRITANI1983}

J.~Iritani, K.~Kuga:
{\em Duality between the Lorenz curves and the income distribution functions},
Economic Studies Quarterly {\bf 34} (1983) 9-21

\bibitem{THISTLE1989}
%	{\color{blue} THISTLE1989}

P.~D.~Thistle:
{\em Duality between generalized Lorenz curves and distribution functions},
Economic Studies Quarterly {\bf 40} (1989) 183-187

\bibitem{AABERGE2000}
%	{\color{blue} AABERGE2000}

R.~Aaberge:
{\em Characterizations of Lorenz curves and income distributions},
Social Choice and Welfare {\bf 17} (2000) 639-653


\bibitem{LORENZCURVE2005}
%	{\color{blue} LORENZCURVE2005}

Christian Kleiber:
{\em The Lorenz curve in economics and econometrics},
Gini-Lorenz Centennial Conference, Siena, May 23-26, 2005, and in 
Gianni Betti and Achille Lemmi (eds.): Advances on Income Inequality and Concentration Measures. Collected Papers in Memory of Corrado Gini and Max O. Lorenz (London Routledge, 2008)
% modern description of the Lorenz curve, centennary


\bibitem{GROWTHRESET}
%	{\color{blue} GROWTHRESET}

T.~S.~B\'{\i}r\'o, and Z.~N\'eda,  {\em Unidirectional random growth with resetting}, 
Physica A: Statistical Mechanics and its Applications (2018), Vol. 499, 335-361


\bibitem{DOGUM1977}
%	{\color{blue} DOGUM1977}

C.~Dogum:
{\em A new model of personal income distributiuons: Specification and estimation},
Economie Appliqu\'ee {\bf 30} (1977) 413-437

\bibitem{TSALLIS1988}

C.~Tsallis:
{\em Possible generalization of Boltzmann-Gibbs statistics}, Journal of Statistical Physics {\bf 52} (1988)
479-487


\bibitem{MCDONALD1984}
%	{\color{blue} MCDONALD1984}

J.~B.~McDonald:
{\em Some generalized functions for the size distribution of income},
Econometrica {\bf 52} (1984) 647-663

\bibitem{TAILLIE1981}
%	{\color{blue} TAILLIE1981}
.~Taillie:
{\em Lorenz ordering within the generalized gamma family of income distributions},
Statistical Distributions in Scientific Work {\bf 6} (1981) 181-192

\bibitem{WILFLING1993}
%	{\color{blue} WILFLING1993}

B.~Wilfling, W.~Kr\"amer:
{\em Lorenz ordering of Singh-Maddala income distributions},
Economic Letters {\bf 43} (1993) 53-57

\bibitem{BOUCHAUD2000}
%	{\color{blue} BOUCHAUD2000}

J.~P.~Bouchaud and M.~Mezard: 
{\em Wealth condensation in a simple model of economy}, 
 Physica A: Statistical Mechanics and its Applications {\bf 282} (2000) 536-545

\bibitem{KLEIBER2002}
%	{\color{blue} KLEIBER2002}

C.~Kleiber, S.~Kotz:
{\em A characterization of income distributions in terms of generalized
Gini coefficients},
Social Choice and Welfare {\bf 19} (2002) 789-794

\bibitem{YITZHAKI1983}
%	{\color{blue} YITZHAKI1983}

S.~Yitzhaki:
{\em On an extension of the Gini inequality index},
International Economic Review {\bf 24} (1983) 617-628

\end{thebibliography}
\end{document}